# Simple metal binary phases based on the body centered cubic structure: electronic origin of distortions and superlattices

**Valentina F Degtyareva and Nataliya S Afonikova**

Institute of Solid State Physics, Russian Academy of Sciences,
Chernogolovka 142432, Russia

E-mail: degtyar@issp.ac.ru

**Abstract.** Binary alloy phases of the noble metals with main group elements are analyzed in relation to body centered cubic structure with orthorhombic and hexagonal distortions. Stability of these distorted phases is considered on the base of the Fermi sphere – Brillouin zone interaction to understand the important part of the band structure energy contribution to the overall crystal energy. Examination of Brillouin zone – Fermi sphere configurations for several representative phases has shown significance of the electron energy contribution in the formation of distorted structures with superlattices and ordered vacancies. This approach may be useful for understanding recently found complex structures in compressed simple alkali metals.

## 1. Introduction

Binary phases of simple metals form an area of exploration for physical–chemical factors of structural stability. The interest to this problem has been renewed because of the recent discoveries of complex structures in elemental metals under pressure (see reviews [1,2] and references therein). Recently very complex structures were observed in compressed light alkalis Li and Na [3-5] considered as "pseudobinary compounds" or "electrides" (see also recent review [6]). Moreover, in compressed sodium [3] and potassium [7,8] the structure of NiAs-$hP$4 type was found as well we its orthorhombic distortion, the $oP$8 structure, both being very common among binary alloy phases. These relations of the elemental structures to binary phases imply some similarities in physical origin of structural complexity. In this paper, among others the structural distortions of the NiAs-type structure are considered in order to gain some insights into the factors of their structural stability.

Crystal structure stability of metallic phases is defined by two main contributions: the electrostatic (or Madelung) energy and electronic band structure energy. The former prefers high-symmetry close-packed structures such as face centered cubic (fcc), close-packed hexagonal (hcp) and body centered cubic (bcc). The band structure energy term depends on valence electron energy which is reduced by the formation of an energy gap on the Brillouin planes near the Fermi level [9,10]. Within the nearly-free electron model the energy gap is formed with the condition when the Fermi sphere radius, $k_F$, is close to the Brillouin plane corresponding to the wave vector $q_{hkl}$, so that $k_F = \frac{1}{2} q_{hkl}$.

Significance of the electron energy contribution into the structure stability is displayed by the family of Hume-Rothery phases in the Cu-Zn and related alloys [11]. The crystal structure of the phases is defined by the number of valence electrons per atom or electron concentration:

fcc (1.35) → bcc (1.5) → complex cubic (1.62) → hcp (1.75).

Simple metal binary phases based on bcc

Along with symmetrical close-packed structures fcc, bcc and hcp there is a complex cubic γ-phase with 52 atoms in the unit cell – a classical example of a complex Hume-Rothery phase. This structure is considered as a 3×3×3 supercell of bcc with 2 atoms removed from the resulting 54-atom cell. In the case of the γ-phase one can see how the band structure term gains over the electrostatic term with the formation of additional Brillouin zone planes close to the Fermi sphere (36 in γ-phase instead of 12 in bcc) and with a better filling of the Brillouin-Jones zone by electronic states (0.93 in γ-phase instead of 0.75 in bcc).

Alkali metals have at ambient pressure the usual metallic bcc structure. Breaking high symmetry and the formation of complex structures can be attributed to effects of band structure energy that increase on compression [12]. "Pressure induced Jones zone activation" is considered theoretically in [13]. The first complex phase in Li and Na with the $cI$16 structure is a supercell of bcc 2×2×2 and slight atomic shifts and appears a consequence of Hume-Rothery effects displaying the formation of new Brillouin planes of (211) type close to the Fermi sphere. This case is quite similar to formation of the γ-phase $Cu_5Zn_8$-$cI$52. Further complex structures of alkalis on pressure increase can be related to the Hume-Rothery mechanism assuming an electron transition as was suggested for sodium for $oP$8 structure, which is structurally identical to some of binary phases as for example to AuGa-$oP$8 with 2 electrons per atom[14]. Therefore analysis of structural relations for binary phases and mechanisms of distortion from high symmetry metallic phases may help in understanding the origin in complexity of compressed elements.

It should be noted that the relationship of various complex alloy structures to the simple metallic structures of bcc, fcc and hcp has been analyzed with analysis of atomic environment types by Sluiter [15]. It is interesting to explore structural relationship by examining for related phases of appropriate Brillouin-Jones zone and corresponding Fermi spheres.

In this paper we consider some representative examples of complex binary phases derived from the initial bcc structure. Alloy phase components are selected from simple metallic elements (Cu and Au with addition of Zn, Al, In and Sn), so that the valence electron contribution is defined by the number of the group in the Periodic Table. An attempt is made to consider some complex binary phases in relation to bcc – one of the basic metallic structures – and the effects of ordering, vacancies and superstructures which lead to minimization of electronic energy. The latter is attained by a closer contact of the Brillouin zone planes with the Fermi sphere and by increase of the number of Brillouin zone planes contacting with the Fermi sphere.

## 2. Method of analysis

Crystal energy contribution due to the band structure energy is defined by configurations of Brillouin – Jones zone planes (called further as BZ) and the Fermi sphere (FS) in the nearly free-electron model. A special program BRIZ has been developed to construct BZ-FS configurations and to calculate some parameters as the Fermi sphere radius ($k_F$), values of reciprocal wave vectors of BZ planes ($q_{hkl}$), volumes of BZ and FS [16]. The BZ planes are selected from the condition $q_{hkl} = 2 k_F$ that have a non-zero structure factor. In this case an energy gap is opened on the BZ plane leading to lowering of the electron energy. The ratio of ½$q_{hkl}$ to $k_F$ is usually less than 1 and equal ~0.95, called as "truncation" factor [17]. On the BZ-FS presentations by the BRIZ program BZ planes cross FS, whereas in real case the Fermi sphere is deformed remaining inside BZ. The "truncation" factor is a characteristic value of a decrease in the electron energy on the BZ plane.

The structure of a phase for the analysis by the BRIZ program is characterized with lattice parameters and the number of atoms in the unit cell, what gives the atomic volume ($V_{at}$). The valence electron concentration (z) is the average number of valence electrons per atom that gives the value of the Fermi sphere radius $k_F = (3\pi z /V_{at})^{1/3}$. Further structure characterization parameters are the number of BZ planes in contact with the FS, the degree of "truncation" factor and the value of BZ filling by electronic

Simple metal binary phases based on bcc

states, defined as a ratio of the volumes of FS and BZ. It should be noted that in the current model the FS volume is a measure of valence electron numbers participating in the band structure contribution even the FS is deformed from a sphere in the real case. As was discussed by Heine [18], the Brillouin-Jones zone that is well accommodated by a FS is called as "Jones zone" ("Jones barrel"); and one of the structure stability conditions is that "the Jones zone being completely filled".

This condition of a BZ almost filled with electronic states has one important consequence – similar crystal structures should contain in their BZ a nearly constant number of valence electrons, zN = const, where N is number of atoms in the unit cell [10]. To satisfy this conditions some vacant structures can be formed as will be discussed further on example of $Cu_3Zn$-$hP$3 phase. Therefore the site occupation is considered here as an important structural characteristic.

All structural data for binary phases considered in this paper have been found in the Pauling File and the standardized crystallographic data are used [19].

## 3. Results and discussion

A representative system with Hume-Rothery phases is the Cu-Zn alloy system consists of two simple metals that are close neighbors in the Periodic Table. In this case, the difference in the chemical properties of constituent elements (atomic radius, electronegativity etc.) is minimal and the phase sequence is defined primary by the electron concentration. There is also minimal effect of ordering and the main Cu-Zn phases are solid solutions based on typical metallic structures fcc, bcc and hcp.

Going from the Cu-Zn system to other related systems additional effects to phase stability should be taken into account leading to more complex structures of ordered type. Behind this complexity, Hume-Rothery effects can be recognized once the Brillouin-Jones zone is constructed that is well accommodating by the FS.

In this paper we consider some binary phases related to bcc with two kinds of distortions: (1) orthorhombic and (2) hexagonal and further formation of superlattices of monoclinic or even triclinic type. For the non-orthogonal cells the BZ presentation is given with related orthogonal axes.

*3.1. Orthorhombic distortions of the bcc structure*

A representative example of orthorhombic displacements on the basis of bcc or CsCl type structure (Pearson symbol $cP$2) is displayed by the AuCd alloy on the decrease of temperature. On alloy cooling a transformation is observed from AuCd-$cP$2 into an AuCd-$oP$4 structure showing an orthorhombic distortion resulting in lattice parameters $a \approx a_{cub}\sqrt{2}$, $b \approx a_{cub}$, $c \approx a_{cub}\sqrt{2}$ (see Table 1).

It is remarkable that the AuCd-$cP$2 phase is isoelectronic to the CuZn-$cI$2 phase (with z = 1.5) which is a classical Hume-Rothery phase with the bcc structure. At low temperature CuZn has an ordered CsCl type structure. Thus there is a similarity between CuZn and AuCd in the common CsCl-type cubic structure, however there is a difference in occurring of the orthorhombic phase $oP$4 in AuCd. Comparing the BZ configurations for AuCd in the cubic and the orthorhombic structures (Figure 1 a,b), one can see that for the $oP$4 phase some of the BZ planes are in a better contact with the FS, as is presented by ratios $k_F/(½q_{hkl})$ in the Table 1. Also the degree of BZ filling by valence electrons is higher for $oP$4 comparing to $cP$2. These data indicate there is a gain in electronic energy for the orthorhombic phase over the cubic one.

At the alloy composition close to AuCd but with some deficiency in Cd there is a monoclinic phase AuCd-$mP$6 (composition $Au_{1.1}Cd_{0.9}$). The structure of this phase is slight distortion of the orthorhombic phase with the lattice parameters $a \approx a_{ort}$, $b \approx b_{ort}$, $c \approx 3/2\ c_{ort}\sin\beta$. This monoclinic distortion is a response of the structural arrangement to a slight decrease in z and $k_F$ in order to improve the BZ-FS interaction with a formation of new BZ planes. Comparing BZ-FS configurations for AuCd-$oP$4 and AuCd-$mP$6

Simple metal binary phases based on bcc

(Figure 1 b,c) one can see that the (200) plane of *oP*4 (in the front) is transformed into the (20$\bar{1}$) plane of *mP*6 with an addition of the (200) plane that is also in contact with FS. Thus for AuCd-*mP*6 there are lower values of $k_F$ and the BZ filling comparing to AuCd-*oP*4, however and some new BZ planes are formed close to FS.

In the Au-Al alloy system there are two orthorhombic phases near the composition of Au$_2$Al derived from bcc or CsCl with structures *oP*12 and *oP*30, the latter is enriched by Al (Table 1). Relation of lattice parameters to the cubic *cP*2 phase is a ≈ 2$a_{cub}$, b ≈ $a_{cub}$, c ≈ 3$a_{cub}$ for *oP*12 and a ≈ 3$a_{cub}$, b ≈ 5$a_{cub}$, c ≈ $a_{cub}$ for *oP*30. These two structures differ by one parameter equal to 2$a_{cub}$ and 5$a_{cub}$ and they are superstructures to a high-temperature phase Au$_2$Al-*tI*6. For these two phases, configurations of BZ and inscribed FS (shown in Figure 1 d, e) are similar to each other and are characterized by the formation of additional BZ planes due to superlattice formation along the a* vector for *oP*12 and along b* for *oP*30 (the b* axis for *oP*30 is shown out of the page to emphasize the similarity with *oP*12).

*3.2. Hexagonal distortions of bcc with vacancies and superlattices*

An example of a hexagonal structure based on bcc is a so-called omega-phase in Ti-group metals formed by alloying or at high pressure [20]. This structure (*hP*3) has lattice parameters related to bcc: a ≈ $a_{cub}\sqrt{2}$ and c ≈ $a_{cub}\sqrt{3}$/2 with the ideal value of c/a = 0.612. One of three atoms is at the cell corner and two atoms occupy trigonal prisms at the height of ~1/2.

In binary phases there is a large family of ordered hexagonal phases with the doubled c axis of compositions A$_2$B and AB represented by structure types B8$_2$-Ni$_2$In (*hP*6) and B8$_1$-NiAs (*hP*4). The B8 structure type is very common in the binary alloy systems and many related phases exist with different types of ordering, vacancies and superlattices. A survey of superstructures in NiAs – Ni$_2$In type phases is given by Lidin [21].

In the present paper we consider some phases related to the B8 type in the binary systems of noble metals and an element from the main groups of the Periodic Table. This allows us to define the phases by a certain value of electron concentration and to apply Hume-Rothery arguments to the stability of these complex phases.

The CuZn$_3$-*hP*3 phase is one of the Hume-Rothery Cu-Zn phases; however it is usually omitted from the discussion of phase sequences along the electron concentration. Atomic positions in *hP*3 are closely related to that of bcc if the latter described using a hexagonal cell. Atoms in trigonal prisms move from heights 1/3 and 2/3 to ~1/2 and these displacements result in appearance of new diffraction peaks. As a consequence, BZ of *hP*3 shown in Figure 2a consists of 18 planes of (110) and (101) types instead of 12 planes of (110) type for bcc.

Another interesting occurrence in the CuZn$_3$-*hP*3 phase [22] is a partial occupation ~0.7 of the Cu site leading the number of atoms in the cell N = 2.7. Vacancies are randomly distributed over the Cu atomic position. Formation of this vacant structure is the result of satisfying the condition that the number of valence electrons in the BZ in a particular phase should be nearly constant, i.q. zN ≈ const, where N is a number of atoms a unit cell.

Thus, the CuZn$_3$-*hP*3 phase is a quite striking example of a structural modification that following the Hume-Rothery mechanism exhibits the appearance of new BZ planes near FS and of the vacancies in atomic positions. Structures of the B8$_2$-*hP*6 type exist in high temperature Cu$_{1.5}$Al and Cu$_2$In phases with partial occupation of Cu atoms in trigonal prisms equal to 0.7 and 0.78, respectively. The formation of vacancies results in a decrease of the number of atoms in the unit cell to 5.4 and 5.56, respectively. These values correlate with the valence electron concentration z of 1.78 and 1.72 giving for both phases nearly equal value zN (~ 9.6-9.7).

Vacancies, disordered at high temperature, have a tendency to order at low temperature and, together with the ordering of constituents, cause a grate variety of superlattice structures of low

Simple metal binary phases based on bcc

symmetry. Some examples of these phases are considered here by analyzing the BZ-FS configurations that make evident the relations of their structures to a basic hexagonal structure of B8-type as shown in Figure 2. Diffraction patterns of the phases consist of groups of strong reflections situated near the location of the (110) and (101) reflections of $hP3$ (Figure 2a). Only the strongest peaks are selected for BZ construction to keep clear the structural relation.

The phase $Au_7In_3$ has a hexagonal crystal structure with 60 atoms in the unit cell related to high temperature phase $Au_9In_4$, as was discussed by Schubert [23]. The phase $Au_9In_4$ has a cubic ($cP52$) structure known as γ-brass, $Cu_8Zn_5$-type. Lattice parameters for $Au_7In_3$ are related to the cubic γ-phase as $a \approx a_\gamma \sqrt{3/2}$, $c \approx a_\gamma \sqrt{3}/2$ or to the hexagonal ω-phase $a \approx 3/2\, a_\omega \sqrt{3}$, $c \approx 3c_\omega$. The $hP60$ structure contains ~20 ω-type cells with 3 atoms in the cell. As discussed by Schubert [16], there are vacancies in the $Au_7In_3$ structure and the distribution of vacancies and atomic displacements are different from that in the high temperature γ-brass type $Au_9In_4$ phase.

The BZ configuration for $Au_7In_3$ shown in Figure 2b only includes planes corresponding to the strongest reflections in the diffraction pattern. If all of the less intense peaks are included that lie close to $2k_F$ the BZ polyhedron will be more multi-faceted and complex and the BZ filling by electronic states will increase (to 0.93).

The monoclinic CuAl phase (Figure 2c) is related to the hexagonal B8 type structure as $a \approx 2c_h \sin\beta$, $b \approx a_h$, $c \approx a_h \sqrt{3}$, where "h" is referred to the B8 cell. The CuAl-$mC20$ phase contains 4 cells of B8 type with the number of atoms in the cell equal $N_{at} = 20/4 = 5$.

Another monoclinic phase $Cu_5Sn_4$ is a superlattice of the B8 structure with relations of lattice vectors: $\mathbf{a} = 2\mathbf{a_h} - \mathbf{c_h}$, $\mathbf{b} = \mathbf{a_{1h}} - \mathbf{a_{2h}}$, ($b = a_h \sqrt{3}$), $\mathbf{c} = 2\mathbf{a_h} + \mathbf{c_h}$, as was considered in [21]. The $Cu_5Sn_4$-$mP36$ phase contains 8 cells of B8 type with the number of atoms in the unit cell equal to $N_{at.} = 36/8 = 4.5$. Comparing this value with that of CuAl-$mC20$, one can see that by increasing the valence electron concentration the number of atoms in the cell decreases to keep zN nearly constant: 2×5 = 10 for CuAl-$mC20$ and 2.3×4.5 = 10.35.

The BZ-FS configurations for both monoclinic phases are shown in Figure 2 c,d displaying close relations of these phases to $CuZn_3$-$hP3$ phase (Figure 2a) and as consequence to the bcc structure. For all three phases we have similar values of BZ filling and ratios $k_F/(½q_{hkl})$ (see Table 2).

One more complex example of a triclinic superstructure ($aP40$) based on the B8 type is the $Cu_7In_3$ phase with lattice parameters $a \approx a_h \sqrt{3}$, $b \approx 2a_h \sin\beta$, $c \approx 2c_h$ (Figure 2c). Interestingly, for this alloy composition high temperature stabilizes the γ-brass type phase $Cu_9In_4$-$cP52$. Whereas in the Cu-Zn system the γ-brass $Cu_8Zn_5$ phase is stable in the whole temperature range, there are several systems that exhibit transitions from the high-temperature γ-brass phase to a low-temperature phase with other structure types: $Au_7In_3$-$hP60$ and $Cu_7In_3$-$aP40$, both considered in this paper. While the γ-brass type structure is referred usually as a pure Hume-Rothery phase stabilized due to the electronic contribution, some other factors in the case of phases in Au-In and Cu-In systems should be taken into account that originate from the chemical difference of components. Nevertheless the Hume-Rothery effects to phase stability are presented for all considered phases, as can be seen from the BZ-FS configurations and from the data obtained from this consideration.

## 4. Conclusion

Structural distortions of the body centered cubic lattice are considered with the formation of orthorhombic, hexagonal, monoclinic and triclinic structures in binary phases of the noble metals and main group elements (Cd, Al, In and Sn). Construction of the Brillouin-Jones polyhedrons by using the planes corresponding to the diffraction peaks with the wave vectors close to $2k_F$ has demonstrated the Hume-Rothery mechanism as the main effect that controls alloy phase stability. In such construction,

Simple metal binary phases based on bcc

there are many BZ planes in contact with the FS that gives a high value of BZ filling by electronic states. Hexagonal distortions of bcc give rise to the existence of many binary phases with superstructures related to the B8 type structure. These structural relations are distinctly shown with the examination of the constructed BZ-FS configurations.

The suggested model of the FS-BZ interaction helps us realize the mechanism and physical origin of the formation of complex superstructures based on a simple metallic bcc structure. This approach will be useful in understanding mechanism of formation and structural relationship of complex structures recently found in elements under high pressure. Unlike in a binary phase, the factor of atomic chemical difference is absent in a pure element and factors of electronic band structure energy become more apparent especially as they increase under pressure. It is remarkable that the structure of NiAs-type, distortions of which are considered for binary phases in this paper, was observed in alkalis Na and K under pressure [7,8]. Along with NiAs-$hP4$ in these elements also an orthorhombic distortion of $oP8$ type was found that can be stabilized by Hume-Rothery effects as considered in [14]. Great variety of complex structures in alkali and alkali-earth metals under pressure need further understanding in physical origin and relation to basic metallic structures.

**Acknowledgment**

The authors gratefully acknowledge Dr. Olga Degtyareva for valuable discussion and comments.

Simple metal binary phases based on bcc

**Table 1.** Structure parameters of several representative Hume-Rothery phases in binary systems based on group-I elements. Pearson symbol, space group and lattice parameters are from literature data. Fermi sphere radius $k_F$, ratios of $k_F$ to distances of Brillouin zone planes ½ $q_{hkl}$ and the filling degree of Brillouin zones by electron states $V_{FS}/V_{BZ}$ are calculated by the program BRIZ.

| Phase | AuCd ht | AuCd | Au$_{1.1}$Cd$_{0.9}$ | Au$_2$Al | Au$_2$Al (Al+) |
|---|---|---|---|---|---|
| Structural data [a] | | | | | |
| Pearson symbol | cP2 | oP4 | mP6 | oP12 | oP30 |
| Space group | $Pm\bar{3}m$ | $Pmma$ | $P12/m1$ | $Pnma$ | $Pnnm$ |
| lattice parameters (Å) | $a$ = 3.324 | $a$ = 4.766<br>$b$ = 3.151<br>$c$ = 4.859 | $a$ = 4.910<br>$b$ = 3.089<br>$c$ = 7.431<br>$\beta$ = 105.38° | $a$ = 6.715<br>$b$ = 3.219<br>$c$ = 8.815 | $a$ = 8.801<br>$b$ = 16.772<br>$c$ = 3.219 |
| FS – BZ data from the BRIZ program | | | | | |
| z (number of valence electrons per atom) | 1.5 | 1.5 | 1.45 | 1.67 | 1.69 |
| $k_F$ (Å$^{-1}$) | 1.342 | 1.345 | 1.333 | 1.460 | 1.467 |
| Total number BZ planes | 12 | 16 | 18 | 18 | 22 |
| $k_F/($½ $q_{hkl})$ max<br>min | 1.004 | 1.040<br>0.927 | 1.0395<br>0.886 | 1.080<br>1.008 | 1.085<br>0.983 |
| $V_{FS}/V_{BZ}$ | 0.750 | 0.768 | 0.751 | 0.891 | 0.889 |

[a] Ref. [19]

Simple metal binary phases based on bcc

**Table 2.** Structure parameters of several representative Hume-Rothery phases in binary systems based on group-I elements. Pearson symbol, space group and lattice parameters are from literature data. Fermi sphere radius $k_F$, ratios of $k_F$ to distances of Brillouin zone planes ½ $q_{hkl}$ and the filling degree of Brillouin zones by electron states $V_{FS}/V_{BZ}$ are calculated by the program BRIZ.

| Phase | $CuZn_3$ | $Au_7In_3$ | CuAl | $Cu_5Sn_4$ | $Cu_7In_3$ |
|---|---|---|---|---|---|
| Structural data [a] | | | | | |
| Pearson symbol | hP3 | hP60 | mC20 | mP36 | aP40 |
| Space group | $P\bar{6}m2$ | $P\bar{3}$ | $C2/m$ | $P12_1/c1$ | $P\bar{1}$ |
| lattice parameters (Å) | $a = 4.275$<br>$c = 2.590$ | $a = 12.215$<br>$c = 8.509$ | $a = 12.065$<br>$b = 4.105$<br>$c = 6.913$<br>$\beta = 124.95°$ | $a = 9.84$<br>$b = 7.27$<br>$c = 9.84$<br>$\beta = 117.5°$ | $a = 6.724$<br>$b = 9.126$<br>$c = 10.071$<br>$\alpha = 73.2°$<br>$\beta = 82.8°$<br>$\gamma = 89.8°$ |
| FS – BZ data from the BRIZ program | | | | | |
| z (number of valence electrons per atom) | 1.73 | 1.62 | 2.0 | 2.3 | 1.61 |
| $k_F$ (Å$^{-1}$) | 1.553 | 1.378 | 1.616 | 1.578 | 1.481 |
| Total number BZ planes | 18 | 18 | 18 | 18 | 18 |
| $k_F/(½\ q_{hkl})$ max<br>min | 1.057<br>1.049 | 1.015<br>1.012 | 1.056<br>1.034 | 1.056<br>1.049 | 1.019<br>1.001 |
| $V_{FS}/V_{BZ}$ | 0.942 | 0.840 | 0.909 | 0.939 | 0.835 |

[a] Ref. [19]

Simple metal binary phases based on bcc

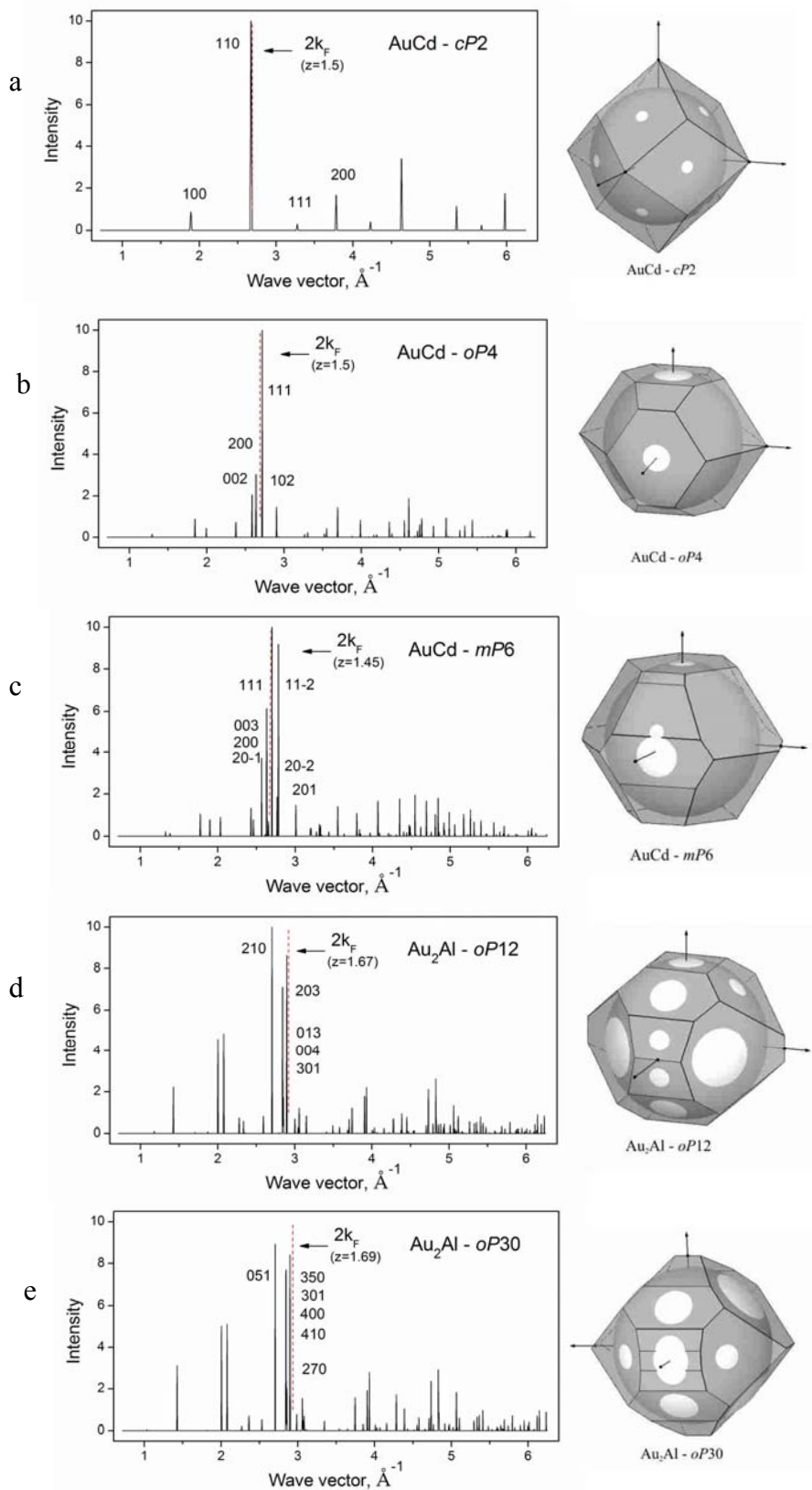

**Figure 1.** Calculated diffraction patterns for selected phases from Table 1 (left) and corresponding Brillouin-Jones zones with the inscribed Fermi spheres (right). The position of $2k_F$ and the hkl indices of the planes used for the BZ construction are indicated on the diffraction patterns.

Simple metal binary phases based on bcc

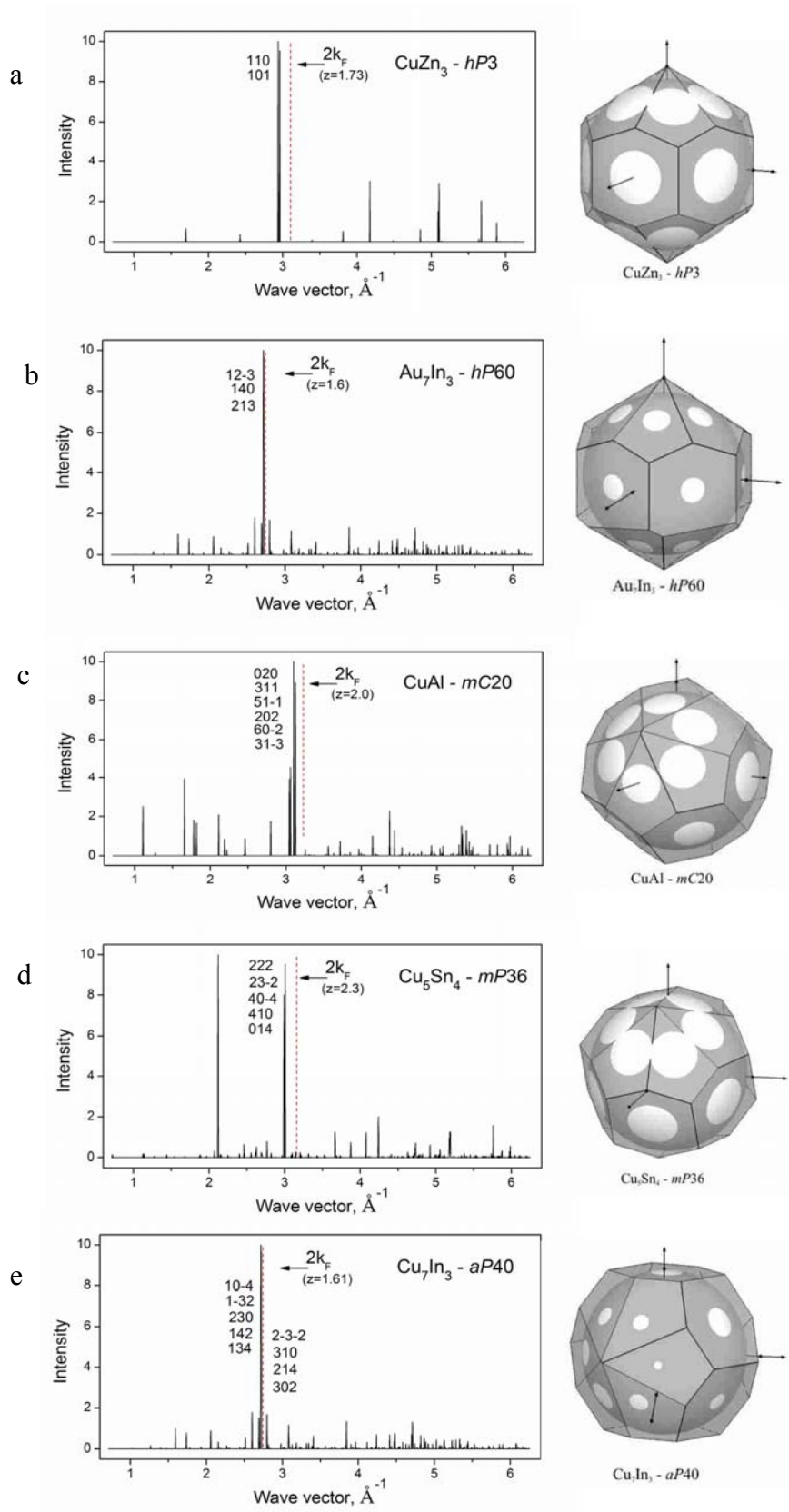

**Figure 2.** Calculated diffraction patterns for selected phases from Table 2 (left) and corresponding Brillouin-Jones zones with the inscribed Fermi spheres (right). The position of $2k_F$ and the hkl indices of the planes used for the BZ construction are indicated on the diffraction patterns.